\documentclass[12pt,showpacs,showkeys,amsmath,amssymb]{revtex4}
\usepackage{amsmath,amsfonts,amsthm,amscd,amssymb,latexsym}
\usepackage{bm}
\usepackage{dcolumn}
\usepackage{graphicx}
\usepackage{epstopdf}
\usepackage{color}
\usepackage{epsf}
\usepackage{epsfig}
\usepackage{graphicx, epic, eepic, color}

\def\@{\partial_}

\def\negenspace{\kern-1.1em}

\def\sqr#1#2{{\vcenter{\hrule height.#2pt\hbox{\vrule width.#2pt
height#1pt \kern#1pt \vrule width.#2pt}\hrule height.#2pt}}}

\date{\today}

\begin{document}

\title{Virial Theorem in Nonlocal Newtonian Gravity}

\author{B. Mashhoon}
\email{mashhoonb@missouri.edu}
\affiliation{Department of Physics and Astronomy,
University of Missouri, Columbia, Missouri 65211, USA}

\begin{abstract} 

Nonlocal gravity is the recent classical nonlocal generalization of Einstein's theory of gravitation in which the past history of the gravitational field is taken into account.   In this theory,  nonlocality appears to  simulate dark matter. The virial theorem for the Newtonian regime of nonlocal gravity theory is derived and its consequences   for ``isolated" astronomical systems in virial equilibrium at the present epoch are investigated. In particular, for a sufficiently isolated nearby \emph{galaxy} in virial equilibrium, \emph{the galaxy's baryonic diameter $\mathcal{D}_0$---namely, the diameter of the smallest sphere that completely surrounds the baryonic system at the present time---is predicted to be larger than  the effective dark matter fraction $f_{DM}$ times a universal length that is the basic nonlocality length scale} $\lambda_0 \approx 3 \pm 2$ kpc.

\end{abstract}



\maketitle

\section{Introduction}

In the standard theory of relativity, physics is local in the sense that a postulate of locality permeates through the special and general theories of relativity. First, Lorentz invariance is extended in a pointwise manner to actual, namely,  accelerated, observers in Minkowski spacetime. This \emph{hypothesis of locality} is then employed crucially in Einstein's local principle of equivalence to render observers pointwise inertial in a gravitational field~\cite{Ei}.  Field measurements are intrinsically nonlocal, however. To go beyond the locality postulate in Minkowski spacetime, the past history of the accelerated observer must be taken into account. The observer in general  carries the memory of its past acceleration. The deep connection between inertia and gravitation suggests that gravity could be nonlocal as well and in nonlocal gravity the gravitational memory of past events must then be taken into account. Along this line of thought, a classical nonlocal generalization of Einstein's theory of gravitation has recently been developed~\cite{NL1, NL2, NL3, NL4, NL5, CM, Mash1, NL6, NL7, Mash2, NL8, NL9}. In this theory the gravitational field is local, but satisfies partial integro-differential field equations. Moreover, a significant observational consequence of this theory is that the nonlocal aspect of gravity appears to simulate dark matter. The physical foundations of this classical theory, from nonlocal special relativity theory to nonlocal general relativity, sets it completely apart from purely phenomenological and \emph{ad hoc} approaches to the problem of dark matter. 

Dark matter is currently required in astrophysics for explaining the gravitational dynamics of galaxies as well as clusters of galaxies~\cite{NL6}, gravitational lensing observations~\cite{NL7} and structure formation in cosmology~\cite{NL9}. We emphasize that only some of the implications of nonlocal gravity theory have thus far been confronted with observation~\cite{NL6, NL8}. It is also important to mention here that many other approaches to nonlocal gravitation theory exist that are, however, inspired by developments in quantum field theory. The consideration of such theories is well beyond the scope of this purely classical work. 

In this paper, we are concerned with the Newtonian regime of nonlocal gravity, where Poisson's equation of Newtonian gravity is modified by the addition of a certain average over the gravitational field. This nonlocal term involves a kernel function $q$ whose functional form can perhaps be derived from a future more complete theory, but, at the present stage of the development of nonlocal gravity, must be determined using observational data. It is necessary that a unique kernel be eventually chosen in this way, but kernel $q$ at the present time could be either $q_1$ or $q_2$~\cite{NL5}. Each of these kernels is spherically symmetric in space and contains three length scales $a_0$, $\lambda_0$, and $\mu_0^{-1}$ such that $a_0<\lambda_0<\mu_0^{-1}$. The basic scale of nonlocality is a galactic length $\lambda_0$ of order 1 kpc, while $a_0$ is a short-range parameter that controls the behavior of $q(r)$ as $r \to 0$. At the other extreme, $r \to \infty$, $q(r)$ decays exponentially as $\exp (-\mu_0 \, r)$, indicating the fading of spatial memory with distance. The short-range parameter $a_0$ is necessary in dealing with the gravitational physics of the Solar System, globular clusters and isolated dwarf galaxies; however, it may be safely neglected in dealing with larger systems such as clusters of galaxies. When $a_0=0$, $q_1$ and $q_2$ reduce to a single kernel $q_0$, $q_1=q_2=q_0$, and the remaining parameters ($\lambda_0$ and $\mu_0$) have been determined from a comparison of the theory with the astronomical data regarding a sample of 12 spiral galaxies from the THINGS catalog---see Ref.~\cite{NL6} for a detailed treatment. The results can be expressed, for the sake of convenience,  as $\lambda_0 \approx 3$ kpc and $\mu_0^{-1} \approx 17$ kpc. Moreover, lower limits have been placed on $a_0$ from the study of the precession of perihelia of planetary orbits in the Solar System~\cite{Ior, DX, NL8}. 

It is interesting to explore the implications of the virial theorem for nonlocal gravity. In general, the virial theorem of Newtonian physics establishes a simple linear relation between the time averages of the kinetic and potential energies of an isolated material system for which the potential energy is a homogeneous function of spatial coordinates.  For an isolated \emph{gravitational} N-body system, the significance of the virial theorem has to do with the circumstance that the kinetic energy is a sum of terms each proportional to the mass of a body in the system, while the potential energy is a sum of terms each proportional to the product of two masses in the system. Thus under favorable conditions, the virial theorem can be used to connect the total dynamic mass of an isolated relaxed gravitational system with its average internal motion.  

The main purpose of the present paper is to discuss, within the Newtonian regime of nonlocal gravity, the consequences of the extension of the virial theorem to nonlocal gravity. Though such an extension is technically straightforward, it is nevertheless physically quite significant as it allows the possibility of making \emph{predictions} regarding the effective dark-matter content of cosmologically nearby isolated N-body gravitational systems in virial equilibrium.

\section{Modification of the Inverse Square Force Law}

It can be shown~\cite{NL8} that in the Newtonian regime of nonlocal gravity,  the force of gravity on point mass $m$ due to point mass $m'$ is given by 
\begin{equation}\label{V1}
\mathbf{F}(\mathbf{r}) = -Gmm'\,\frac{\hat{\mathbf{r}}}{r^2}\,\left\{[1-{\cal E}(r)+ \alpha_0] - \alpha_0\,(1+\frac{1}{2}\,\mu_0\,r)\,e^{-\mu_0\,r}\right\}\,,
\end{equation}
where $\mathbf{r}=\mathbf{x}_{m}-\mathbf{x}_{m'}$, $r=|\mathbf{r}|$ and $\hat{\mathbf{r}}=\mathbf{r}/r$. The quantity in curly brackets is henceforth denoted by $1+ \mathbb{N}$,  where $\mathbb{N}$ is the contribution of nonlocality to the force law and depends upon three parameters, namely, $\alpha_0$, $\mu_0$ and a short-range parameter $a_0$ that is contained in $\mathcal{E}$; in fact, $\mathcal{E}=0$ when $a_0=0$.   We will show in the next section  that $\mathbb{N}$ starts out from zero at $r=0$ with vanishing slope and monotonically increases toward an asymptotic value of about 10 as $r \to \infty$. Thus the gravitational force~\eqref{V1} is \emph{always attractive}; moreover, this force is central, conservative and satisfies Newton's third law of motion. 

Nonlocal gravity is in the early stages of development and depending on whether we choose kernel 
$q_1$ or kernel $q_2$, $\mathcal{E}(r)$ at the present time can be either 
\begin{equation}\label{V2}
 {\cal E}_1(r)=\frac{a_0}{\lambda_0}\,e^{p}\Big[E_1(p)-E_1(p+\mu_0 r)\Big]\,
\end{equation}
or
\begin{equation}\label{V3}
 {\cal E}_2(r)=\frac{a_0}{\lambda_0}\left \{-\frac{r}{r+a_0}e^{-\mu_0 r}+2e^{p} \Big[E_1(p)-E_1(p+\mu_0 r) \Big] \right \}\,,
\end{equation}
respectively, where $p:= \mu_0\,a_0$, $\lambda_0=2/(\alpha_0\, \mu_0)$ and $E_1(u)$ is  the \emph{exponential integral function}~\cite{A+S}
\begin{equation}\label{V4}
E_1(u):=\int_{u}^{\infty}\frac{e^{-t}}{t}dt\,.
\end{equation}
For $u: 0 \to \infty$,  $E_1(u) > 0$ monotonically decreases from infinity to zero. In fact, near $u=0$,  $E_1(u)$  behaves like $-\ln u$ and as $u \to \infty$, $E_1(u)$ vanishes exponentially.  Furthermore, 
\begin{equation}\label{V5}
E_1(x)=-\,C-\ln x -\sum_{n=1}^{\infty}\frac{(-x)^n}{n~ n!}\,,
\end{equation}
where $C=0.577\dots$ is Euler's constant. It is useful to note that 
\begin{equation}\label{V6}
 \frac{e^{-u}}{u+1} < E_1(u) \le \frac{e^{-u}}{u}\,,
\end{equation}
see formula 5.1.19 in Ref.~\cite{A+S}. 

It is clear from Eq.~\eqref{V1} that $\alpha_0$ is dimensionless, while $\mu_0^{-1}$, $\lambda_0$ and $a_0$ have dimensions of length. In fact, we expect that $a_0<\lambda_0<\mu_0^{-1}$; moreover, the short-range parameter $a_0$ and $\mathcal{E}$ may be neglected in Eq.~\eqref{V1} when dealing with 
the rotation curves of spiral galaxies and the internal  gravitational physics of clusters of galaxies. In this way, $\alpha_0$ and $\mu_0$ have been \emph{tentatively} determined from a detailed comparison of nonlocal gravity with observational data~\cite{NL6}: 
\begin{equation}\label{V7}
\alpha_0 = 10.94\pm2.56\,, \qquad  \mu_0 = 0.059\pm0.028~{\rm kpc^{-1}}\,.
\end{equation}
Hence,  we find  $\lambda_0 = 2/(\alpha_0\, \mu_0) \approx 3 \pm 2~ {\rm kpc}$. It is important to mention here that \emph{$\lambda_0$ is the fundamental length scale of nonlocal gravity at the present epoch}; indeed, for $\lambda_0 \to \infty$, $\mathbb{N} \to 0$ and Eq.~\eqref{V1} reduces to Newton's inverse square force law. In what follows, we usually assume that $\alpha_0 \approx 11$ and $\mu_0^{-1} \approx 17$ kpc for the sake of convenience. Furthermore, we expect that $p=\mu_0\,a_0$ is such that $0< p < \frac{1}{5}$. In Ref.~\cite{NL8}, preliminary lower limits have been placed on $a_0$  on the basis of current data regarding planetary orbits in the Solar System. For instance, using the data for the orbit of Saturn, a preliminary lower limit of  $a_0 \gtrsim 2 \times 10^{15}$  cm can be established if we use $\mathcal{E}_1$, while   $a_0 \gtrsim 5.5 \times10^{14}$ cm if we use $\mathcal{E}_2$. 

Let us note that
\begin{equation}\label{V8}
\frac{d\mathcal{E}_1}{dr} = \frac{a_0}{\lambda_0}\, \frac{1}{a_0+r}\,e^{-\mu_0\,r}\,
\end{equation}
and
\begin{equation}\label{V9}
\frac{d\mathcal{E}_2}{dr} = \frac{a_0}{\lambda_0}\,\frac{a_0+2r+\mu_0\,r(a_0+r)}{(a_0+r)^2}\,e^{-\mu_0\,r}\,.
\end{equation}
Therefore, $\mathcal{E}_1(r)$ and $\mathcal{E}_2(r)$ start from zero at $r=0$ and monotonically increase as   $r \to \infty$; furthermore, they asymptotically approach  $\mathcal{E}_1(\infty)=\mathcal{E}_{\infty}$  and $\mathcal{E}_2(\infty)= 2\,\mathcal{E}_{\infty}$, respectively, where 
\begin{equation}\label{V10}
 \mathcal{E}_{\infty}=\frac{1}{2}\,\alpha_0\, p\,e^{p}E_1(p)\,.
\end{equation}
It is a consequence of  Eq.~\eqref{V6} that  $\mathcal{E}_{\infty} < \alpha_0/2$, 
so that in the gravitational force~\eqref{V1}, 
\begin{equation}\label{V11}
\alpha_0 -  \mathcal{E}(r) >0\,.
\end{equation}

In the Newtonian regime, where we formally let the speed of light $c \to \infty$, retardation effects vanish and gravitational memory is purely spatial. The resulting gravitational force~\eqref{V1} thus consists of two parts: an enhanced attractive ``Newtonian" part and a repulsive fading spatial memory (``Yukawa") part with an exponential decay length of $\mu_0^{-1} \approx 17$ kpc. Formula~\eqref{V1} is such that it reduces to Newton's inverse square force  law for $r\to 0$, as it should~\cite{A1, A2, A3, A4, LL}, and on galactic scales, it is a generalization of  the phenomenological Tohline--Kuhn modified gravity approach to the flat rotation curves of spiral galaxies~\cite{T1, T2, K1, K2}. An excellent review of the Tohline--Kuhn work is contained  in  the paper of Bekenstein~\cite{B}. 

For $r \gg \mu_0^{-1}$, the exponentially decaying (``fading memory") part of Eq.~\eqref{V1} can be neglected and
\begin{equation}\label{V12}
\mathbf{F}(\mathbf{r}) \approx -\frac{Gmm'\,[1+ \alpha_0- \mathcal{E}(\infty)] }{r^2}\,\hat{\mathbf{r}}\,,
\end{equation} 
so that $m'\,[\alpha_0-{\cal E}(\infty)]$ has the interpretation of the \emph{total effective dark mass} associated with $m'$. For $a_0=0$, the net effective dark mass associated with point mass $m'$ is simply $\alpha_0\,m'$, where $\alpha_0 \approx 11$~\cite{NL6}. On the other hand, for $a_0 \ne 0$, the corresponding result is $\alpha_0\, \epsilon (p)\, m'$, where
\begin{equation}\label{V13}
\epsilon_1(p) = 1-\frac{1}{2}\,p\,e^p\,E_1(p)\,, \qquad \epsilon_2(p) = 1-p\,e^p\,E_1(p)\,, 
\end{equation}
depending on whether we use $\mathcal{E}_1$ or $\mathcal{E}_2$, respectively. The functions in Eq.~\eqref{V13} start from unity at $p=0$ and decrease monotonically to $\epsilon_1(0.2) \approx 0.85$ and $\epsilon_2(0.2) \approx 0.70$ at $p=0.2$; they are plotted in Figure 1\ of Ref.~\cite{NL8} for $p: 0 \to 0.2$. If $a_0$ turns out to be just a few parsecs or smaller, for instance, then $\epsilon_1\approx \epsilon_2 \approx 1$.

A detailed investigation reveals that it is possible to approximate the exterior gravitational force due to a star or a planet by assuming that its mass is concentrated at its center~\cite{NL8}. In this connection, we note that the radius of a star or a planet is generally much smaller than the length scales $a_0$, $\lambda_0$ and $\mu_0^{-1}$ that appear in the nonlocal contribution to the gravitational force.
Therefore, one can employ Eq.~\eqref{V1} in the approximate treatment of the two-body problem in astronomical systems such as binary pulsars and the Solar System, where possible deviations from general relativity may become measurable in the future. 

Consider, for instance, the deviation from the Newtonian inverse square force law, namely, 
\begin{equation}\label{V14}
\delta\,\mathbf{F}(\mathbf{r}) = -\, \frac{Gmm' \,\hat{\mathbf{r}}}{r^2}\,\mathbb{N}(r)\,.
\end{equation}
For $r<a_0$, it is possible to show via an expansion in powers of $r/a_0$ that~\cite{NL8} 
\begin{equation}\label{V15}
\delta\, \mathbf{F}_1(\mathbf{r}) = -\frac{1}{2} \, \frac{Gmm'}{\lambda_0\,a_0}\,(1+ p)\,\hat{\mathbf{r}}+ \frac{1}{3} \, \frac{Gmm'}{\lambda_0\,a_0}\,(1+ p+ p^2 )\,\frac{r}{a_0}\,\hat{\mathbf{r}}+\cdots\,
\end{equation}
if $\mathcal{E}_1$ is employed, or 
\begin{equation}\label{V16}
\delta\, \mathbf{F}_2(\mathbf{r}) = -\frac{1}{3} \, \frac{Gmm'}{\lambda_0\,a_0}\,(1+ p)\,\frac{r}{a_0}\,\hat{\mathbf{r}}+\cdots\,
\end{equation}
if $\mathcal{E}_2$ is employed. Perhaps dedicated missions, such as ESA's Gaia mission that was launched in 2013, can measure the imprint of nonlocal gravity in the Solar System~\cite{HHP, BDG}. In this connection, we note that 
\begin{equation}\label{V17}
\frac{1}{2} \, \frac{G\,M_{\odot}}{\lambda_0\,a_0}\,(1+ p)\approx \Big(\frac{10^{18}\,{\rm cm}}{a_0}\Big)\,10^{-14}\, {\rm    cm\, s^{-2}}\,,
\end{equation}
which, combined with lower limits on $a_0$ established in Ref.~\cite{NL8}, is at least three orders of magnitude smaller than the acceleration involved in the Pioneer anomaly ($\sim 10^{-7}\, {\rm    cm\, s^{-2}}$). It follows from these results that nonlocal gravity is consistent with the gravitational physics of the Solar System. 

\section{Virial Theorem}

Consider an idealized  isolated system of $N$ Newtonian point particles with fixed masses $m_i$, $i=1,2,\ldots, N$. We assume that the particles occupy a finite region of space and interact with each other only gravitationally such that the center of mass of the isolated system is at rest in a global inertial frame and the isolated system permanently occupies a compact region of space. The equation of motion of the particle with mass $m_i$ and state $(\mathbf{x}_i, \mathbf{v}_i)$ is then
\begin{equation}\label{B1}
m_i\, \frac{d\,\mathbf{v}_i}{dt}= -\sum_{j}{'}~ \frac{G\,m_i\,m_j\,(\mathbf{x}_i-\mathbf{x}_j)}{|\mathbf{x}_i-\mathbf{x}_j|^3}\, [1+ \mathbb{N}(|\mathbf{x}_i-\mathbf{x}_j|)]\,
\end{equation}
for $j=1,2,\ldots, N$, but the case $j=i$ is excluded in the sum by convention. In fact,  a prime over the summation sign indicates that in the sum $j\ne i$. 
Here $1+ \mathbb{N}(r)$ is a \emph{universal} function that is inside the curly brackets in Eq.~\eqref{V1}  and the contribution of nonlocality,  $\mathbb{N}(r)$,  is given by 
\begin{equation}\label{B2}
\mathbb{N}(r) = \alpha_0\,\left[1-(1+\frac{1}{2}\,\mu_0\,r)\,e^{-\mu_0\,r} \right]-\mathcal{E}(r)\,.
\end{equation}

Consider next  the quantities 
\begin{equation}\label{B3}
\mathbb{I}= \frac{1}{2} \sum_i m_i\, x_i^2\,, \qquad \frac{d\,\mathbb{I}}{dt} = \sum_i m_i\, \mathbf{x}_i\cdot \mathbf{v}_i\,,
\end{equation}
where $x_i=|\mathbf{x}_i|$ and 
\begin{equation}\label{B4}
\frac{d^2\,\mathbb{I}}{dt^2}=  \sum_i m_i\, v_i^2 + \sum_i m_i\, \mathbf{x}_i\cdot \frac{d\,\mathbf{v}_i}{dt}\,.
\end{equation}
It follows from Eq.~\eqref{B1} that 
\begin{equation}\label{B5}
\sum_i m_i\,\mathbf{x}_i \cdot \frac{d\,\mathbf{v}_i}{dt}= -\sum_{i, j}{'}~\frac{G\,m_i\,m_j\,(\mathbf{x}_i-\mathbf{x}_j)\cdot \mathbf{x}_i}{|\mathbf{x}_i-\mathbf{x}_j|^3}\, [1+ \mathbb{N}(|\mathbf{x}_i-\mathbf{x}_j|)]\,.
\end{equation}
Exchanging $i$ and $j$ in the expression on the right-hand side of Eq.~\eqref{B5}, we get
\begin{equation}\label{B6}
\sum_i m_i\,\mathbf{x}_i \cdot \frac{d\,\mathbf{v}_i}{dt} = \sum_{i, j}{'}~\frac{G\,m_i\,m_j\,(\mathbf{x}_i-\mathbf{x}_j)\cdot \mathbf{x}_j}{|\mathbf{x}_i-\mathbf{x}_j|^3}\, [1+ \mathbb{N}(|\mathbf{x}_i-\mathbf{x}_j|)]\,.
\end{equation}
Adding Eqs.~\eqref{B5} and ~\eqref{B6} results in 
\begin{equation}\label{B7}
\sum_i m_i\,\mathbf{x}_i \cdot \frac{d\,\mathbf{v}_i}{dt}= -\frac{1}{2}\sum_{i, j}{'}~\frac{G\,m_i\,m_j}{|\mathbf{x}_i-\mathbf{x}_j|}\, [1+ \mathbb{N}(|\mathbf{x}_i-\mathbf{x}_j|)]\,.
\end{equation}
Using this result, Eq.~\eqref{B4} takes the form
\begin{equation}\label{B8}
\frac{d^2\,\mathbb{I}}{dt^2} =  \sum_i m_i\, v_i^2 -\frac{1}{2}\sum_{i, j}{'}~\frac{G\,m_i\,m_j}{|\mathbf{x}_i-\mathbf{x}_j|}\, [1+ \mathbb{N}(|\mathbf{x}_i-\mathbf{x}_j|)]\,.
\end{equation}
Let us recall that the net kinetic energy and the Newtonian gravitational potential energy of the system are given by 
\begin{equation}\label{B9}
\mathbb{T}= \frac{1}{2} \sum_i m_i\, v_i^2 \,, \qquad \mathbb{W}_N=-\frac{1}{2}\sum_{i, j}{'}~\frac{G\,m_i\,m_j}{|\mathbf{x}_i-\mathbf{x}_j|}\,.
\end{equation}
Hence, 
\begin{equation}\label{B10}
\frac{d^2\,\mathbb{I}}{dt^2} = 2\,\mathbb{T} + \mathbb{W}_N + \mathbb{D}\,,
\end{equation}
where
\begin{equation}\label{B11}
\mathbb{D} = - \frac{1}{2}\sum_{i, j}{'}~\frac{G\,m_i\,m_j}{|\mathbf{x}_i-\mathbf{x}_j|}\, \mathbb{N}(|\mathbf{x}_i-\mathbf{x}_j|)\,
\end{equation}
and $\mathbb{N}$ is given by Eq.~\eqref{B2}. 

Finally, we are interested in the average of Eq.~\eqref{B10} over time. Let $< f >$ denote the time average of $f$, where 
\begin{equation}\label{B12}
<f>~ = \lim_{\tau \to \infty} \frac{1}{\tau} \int_0^{\tau} f(t)\,dt \,.
\end{equation}
Then, it follows from averaging Eq.~\eqref{B10} over time that 
\begin{equation}\label{B13}
2\,<\mathbb{T}>~ = -  <\mathbb{W}_N > - <\mathbb{D} > \,,
\end{equation}
since $d\,\mathbb{I}/ dt$, which is the sum of $m\, \mathbf{x}\cdot\mathbf{v}$ over all particles in the system, is a bounded function of time and hence the time average of  $d^2\,\mathbb{I}/ dt^2$ vanishes.  This is clearly based on the assumption that the spatial coordinates and velocities of all particles indeed remain finite for all time. Equation~\eqref{B13} expresses the \emph{virial theorem} in nonlocal Newtonian gravity. 

It is important to digress here and re-examine some of the assumptions involved in our derivation of the virial theorem. In general, any consequence of the gravitational interaction involves the whole mass-energy content of the universe due to the universality of the gravitational interaction; therefore, an astronomical system may be considered isolated only to the extent that the tidal influence of the rest of the universe on the internal dynamics of the system can be neglected.  Moreover, the parameters of the force law~\eqref{V1} refer to the present epoch and hence the virial theorem~\eqref{B13} ignores cosmological evolution. Thus the temporal average over an infinite period of time in Eq.~\eqref{B13} must be reinterpreted here to mean that the relatively isolated system under consideration has evolved under its own gravity such that it is at the present epoch in a steady equilibrium state. That is, the system is currently in virial equilibrium. Finally, we recall that a point particle of mass $m$ in Eq.~\eqref{B13} could reasonably represent a star of mass $m$ as well, where the mass of the star is assumed to be concentrated at its center.  

The deviation of the virial theorem~\eqref{B13} from the Newtonian result is contained in $<\mathbb{D} >$, where $\mathbb{D}$ is given by Eq.~\eqref{B11}. More explicitly, we have 
\begin{equation}\label{B14}
\mathbb{D} = - \frac{1}{2}\sum_{i, j}{'}~\frac{G\,m_i\,m_j}{|\mathbf{x}_i-\mathbf{x}_j|}\, \Big[\alpha_0 - \alpha_0\,(1+\frac{1}{2}\,\mu_0\,|\mathbf{x}_i-\mathbf{x}_j|)\,e^{-\mu_0\,|\mathbf{x}_i-\mathbf{x}_j|} -\mathcal{E}(|\mathbf{x}_i-\mathbf{x}_j|)\Big]\,.
\end{equation}
It proves useful at this point to study some of the properties of the function $\mathbb{N}$, which is  the contribution of nonlocality that is inside the square brackets in Eq.~\eqref{B14}. The argument of this function is $|\mathbf{x}_i-\mathbf{x}_j|>0$ for $i\ne j$; therefore, $|\mathbf{x}_i-\mathbf{x}_j|$ varies over the interval $(0, \mathcal{D}_0]$, where $ \mathcal{D}_0$ is the largest possible distance between any two baryonic point masses in the system. Thus  $\mathbb{N}(r)$, in the context of the virial theorem,  is defined for the interval $0 < r \le \mathcal{D}_0$, where  $\mathcal{D}_0$ is the  diameter of the smallest  sphere that completely encloses the \emph{baryonic} system for all time.  In general, however,  $\mathbb{N}(0) = 0$ and $\mathbb{N}(\infty)=\alpha_0-\mathcal{E}(\infty) >0$, where $\mathcal{E}(\infty) = \mathcal{E}_{\infty}$ or  $2\,\mathcal{E}_{\infty}$,  depending on whether  we use $\mathcal{E}_1$ or $\mathcal{E}_2$, respectively. Moreover,  $d\,\mathbb{N}(r)/dr$ is given by
\begin{equation}\label{B15}
\frac{d}{dr}\,\mathbb{N}_1(r)= \frac{1}{2}\,\alpha_0\,\mu_0\, \frac{r\,[1+\mu_0\,(a_0+r)]}{a_0+r}\,\,e^{-\mu_0\,r}\,,
\end{equation}
if  we use $\mathcal{E}_1$ or
\begin{equation}\label{B16}
\frac{d}{dr}\,\mathbb{N}_2(r) = \frac{1}{2}\,\alpha_0\,\mu_0\, \frac{r^2\,[1+\mu_0\,(a_0+r)]}{(a_0+r)^2}\,\,e^{-\mu_0\,r}\,,
\end{equation}
if  we use $\mathcal{E}_2$. Writing $\exp{(\mu_0\, r)}=1+ \mu_0\,r +\mathcal{R}$, where $\mathcal{R} >0$ represents the remainder of the power series, it is straightforward to see that for $r\ge 0$ and $n=1,2, \dots$,  
\begin{equation}\label{B16a}
e^{\mu_0\,r}\,(a_0+r)^n > r^n\, [1+ \mu_0\,(a_0+r)]\,. 
\end{equation}
This result, for $n=1$ and $n=2$,  implies that the right-hand sides of Eqs.~\eqref{B15} and~\eqref{B16}, respectively, are less than $\alpha_0\,\mu_0/2$. Therefore, it follows that  in general
\begin{equation}\label{B16b}
\frac{d}{dr}\,\mathbb{N}(r)  < \frac{1}{2}\,\alpha_0\,\mu_0\,.
\end{equation}
Moreover, for $r>0$,  Eq.~\eqref{B16b} implies
\begin{equation}\label{B16c}
\mathbb{N}(r) = \int_0^r \frac{d\,\mathbb{N}(x)}{dx} ~dx < \frac{1}{2}\,\alpha_0\,\mu_0\,r.
\end{equation}

We conclude that $\mathbb{N}$ is a monotonically increasing function of $r$ that is zero at $r=0$ with a slope that vanishes at $r=0$.  For  $r \gg \mu_0^{-1}$, $\mathbb{N}(r)$ asymptotically approaches a constant $\alpha_0\, \epsilon:= \alpha_0-\mathcal{E}(\infty)$. Here $\epsilon (p)$ is either $\epsilon_1(p)$ or $\epsilon_2(p)$ depending on whether we use $\mathcal{E}_1$ or $\mathcal{E}_2$, respectively. The functions  $\epsilon_1(p)$ and $\epsilon_2(p)$  are defined in Eq.~\eqref{V13}.

\section{Dark Matter} 

Most of the matter in the universe is currently thought to be in the form of certain elusive particles that have not been directly detected~\cite{NDM1, NDM2, NDM3, NDM4}. The existence and properties of this \emph{dark matter} have thus far been deduced only through its gravity. We are interested here in dark matter only as it pertains to stellar systems such as galaxies and clusters of galaxies~\cite{Zw1, Zw2, RF, RW, SR, Sei, HMK}. We mention that dark matter is also essential in the explanation of gravitational lensing observations~\cite{BC1, BC2} and in the solution of the problem of structure formation in cosmology~\cite{NL9, CCBM}; however, these topics are beyond the scope of this work. 

Actual (mainly baryonic) mass is observationally estimated for astronomical systems using the mass-to-light ratio $M/L$. However, it turns out that the dynamic mass of the system is usually larger and this observational fact is normally attributed to the possible existence of nonbaryonic dark matter. 
Let $M$ be the baryonic mass and $M_{DM}$ be the mass of the nonbaryonic dark matter needed to explain the gravitational dynamics of the system. Then, 
\begin{equation}\label{C1}
f_{DM} := \frac{M_{DM}}{M}  
\end{equation}
is the dark matter fraction and $M+M_{DM}=M\,(1+f_{DM})$ is the dynamic mass of the system. 

In observational astrophysics, the virial theorem of Newtonian gravity is interpreted to be a relationship between the dynamic (virial) mass of the entire system and its average internal motion deduced from the rotation curve or velocity dispersion of the bound collection of masses in virial equilibrium. Therefore, regardless of how the net amount of dark matter in galaxies and clusters of galaxies is operationally  estimated and the corresponding $f_{DM}$ is thereby determined, for sufficiently isolated self-gravitating astronomical systems in virial equilibrium, we must have 
\begin{equation}\label{C2}
2\,<\mathbb{T}>~ = - (1+f_{DM}) <\mathbb{W}_N >\,.
\end{equation}
That is,  virial theorem~\eqref{C2} is employed in astronomy to infer in some way the total dynamic mass of the system. Indeed, Zwicky first noted the need for  dark matter  in his application of the standard virial theorem of Newtonian gravity to the Coma cluster of galaxies~\cite{Zw1, Zw2}.

\section{Effective Dark Matter}

A significant physical consequence of nonlocal gravity theory is that it appears to simulate dark matter~\cite{NL6}. In particular, in the Newtonian regime of nonlocal gravity, the Poisson equation is modified such that the density of ordinary matter $\rho$ is accompanied by a term $\rho_D$ that is obtained from the folding (convolution) of $\rho$ with the reciprocal kernel of nonlocal gravity. Thus $\rho_D$ has the interpretation of the density of \emph{effective dark matter} and $\rho + \rho_D$ is the density of the \emph{effective dynamic mass}.   

The virial theorem makes it possible to elucidate in a simple way the manner in which nonlocality can simulate dark matter.  It follows from a comparison of Eqs.~\eqref{B13} and~\eqref{C2} that nonlocal gravity can account for this ``excess mass" if 
\begin{equation}\label{B18}
<\mathbb{D} >~ = f_{DM} <\mathbb{W}_N >\,,
\end{equation}
where $\mathbb{D}$ and $\mathbb{W}_N$ are given in Eqs.~\eqref{B11} and~\eqref{B9}, respectively. 

It is interesting to apply the virial theorem of nonlocal gravity to sufficiently isolated astronomical N-body systems. The configurations that we briefly consider below consist of clusters of galaxies with diameters $\mathcal{D}_0 \gg \mu_0^{-1} \approx 17$ kpc, galaxies with  $\mathcal{D}_0 \sim \mu_0^{-1}$ and globular star clusters with $\mathcal{D}_0 \ll \mu_0^{-1}$. The results presented in this section follow from certain general properties of the function $\mathbb{N}(r)$ and are completely independent of how the baryonic matter is distributed within the astronomical system under consideration. 

We emphasize that, after setting the short-range parameter $a_0=0$, the parameters $\alpha_0$ and $\mu_0$, and hence $\lambda_0$, were originally determined from the combined observational data for the rotation curves of a sample of 12 nearby spiral galaxies from the THINGS catalog~\cite{NL6}.  These tentative values are given in Eq.~\eqref{V7}. These parameter values were then found to be in reasonable agreement with the internal dynamics of a sample of 10 rich nearby clusters of galaxies from the Chandra x-ray catalog~\cite{NL6}. In the present paper, we use these parameter values to make predictions about \emph{all} nearby isolated N-body gravitational systems that are in virial equilibrium. 

\subsection{Clusters of Galaxies: $f_{DM} \approx \alpha_0\, \epsilon(p)$}

Consider, for example, a cluster of galaxies, where nearly all of the relevant distances are much larger than $\mu_0^{-1}\approx 17$ kpc. In this case, $\mu_0\,r \gg 1$ and hence $\mathbb{N}$ approaches its asymptotic value, namely, 
\begin{equation}\label{B19}
\mathbb{N}  \approx   \alpha_0\, \epsilon(p)\,,
\end{equation}
where $\epsilon = \epsilon_1$ or $ \epsilon_2$, defined in Eq.~\eqref{V13}, depending on whether we use $\mathcal{E}_1$ or $\mathcal{E}_2$, respectively. Hence, Eq.~\eqref{B11} can be written as 
\begin{equation}\label{B20}
<\mathbb{D}> \, \approx  \alpha_0\, \epsilon(p)\,<\mathbb{W}_N >\,.
\end{equation}
It then follows from Eq.~\eqref{B18} that for galaxy clusters
\begin{equation}\label{B21}
f_{DM} \approx \alpha_0\, \epsilon(p)\,
\end{equation}
in nonlocal gravity. We recall that $\epsilon$ is only weakly sensitive to the magnitude of $a_0$.
It follows from $\alpha_0 \approx 11$  that $f_{DM}$ for galaxy clusters is about 10, in general agreement with observational data~\cite{NL6}.  This theoretical result is essentially equivalent to the work on galaxy clusters contained in Ref.~\cite{NL6}, except that Eq.~\eqref{B21} takes into account the existence of the short-range parameter $a_0$. 

\emph{Nonlocal gravity thus predicts that the effective dark matter fraction $f_{DM}$ has approximately the same constant value of about 10 for all isolated nearby clusters of galaxies that are in equilibrium.}

\subsection{Galaxies: $f_{DM} < \mathcal{D}_0/\lambda_0$}
 
Consider next a sufficiently isolated galaxy of diameter $\mathcal{D}_0$ in virial equilibrium. In this case, we recall that $\mathbb{N}(r)$ is a monotonically increasing function of $r$, so that for $0 < r \le \mathcal{D}_0$,  Eq.~\eqref{B16c} implies
\begin{equation}\label{B22}
\mathbb{N}(r) \le  \mathbb{N}(\mathcal{D}_0) < \frac{1}{2}\,\alpha_0\,\mu_0\,\mathcal{D}_0.
\end{equation}
Therefore, it follows from Eq.~\eqref{B11} that in this case
\begin{equation}\label{B23}
\mathbb{D} > (\frac{1}{2}\,\alpha_0\,\mu_0\,\mathcal{D}_0)\, \mathbb{W}_N .
\end{equation}
The virial theorem for nonlocal gravity in the case of an isolated galaxy is then
\begin{equation}\label{B24}
2\,<\mathbb{T}> + <\mathbb{W}_N >~ < \,-\, (\frac{1}{2}\,\alpha_0\,\mu_0\,\mathcal{D}_0)\, <\mathbb{W}_N >\,,
\end{equation}
which means, when compared with Eq.~\eqref{C2}, that  
\begin{equation}\label{B25}
f_{DM} < \frac{1}{2}\,\alpha_0\,\mu_0\,\mathcal{D}_0\,.
\end{equation}
Let us note that 
\begin{equation}\label{B26}
\frac{1}{2}\,\alpha_0\,\mu_0 = \frac{1}{\lambda_0}\,,
\end{equation}
where $\lambda_0$ is the basic nonlocality length scale. Its exact value is not known; however,  from the results of Ref.~\cite{NL6}, we have $\lambda_0 \approx 3\pm2$ kpc. If we formally let $\lambda_0 \to \infty$, then Eq.~\eqref{B25}, namely, $f_{DM} < \mathcal{D}_0/\lambda_0$, implies that in this case nonlocality and the effective dark matter both disappear, as expected. Therefore, for a sufficiently isolated galaxy in virial equilibrium, the ratio of its baryonic diameter  to dark matter fraction $f_{DM}$ must always be above a fixed length $\lambda_0$ of about $3\pm2$ kpc; that is, 
\begin{equation}\label{B27}
\frac{\mathcal{D}_0}{f_{DM}} >  \lambda_0\,.
\end{equation}

To illustrate relation~\eqref{B27}, consider, for instance, the Andromeda Galaxy (M31) with a diameter $\mathcal{D}_0$ of about 67 kpc. In this case,  we have $f_{DM} \approx 12.7$~\cite{Bar1, Bar2}, so that for this spiral galaxy
\begin{equation}\label{B28}
\frac{\mathcal{D}_0}{f_{DM}} \,({\rm Andromeda~Galaxy}) \approx 5.3~ {\rm kpc}\,.
\end{equation}
More recently, the distribution of dark matter in M31~has been further studied in Ref.~\cite{Tam}. Similarly, for the Triangulum Galaxy (M33), we have $\mathcal{D}_0 \approx 34$ kpc and $f_{DM} \approx 5$~\cite{Cor}, so that 
\begin{equation}\label{B29}
\frac{\mathcal{D}_0}{f_{DM}} \,({\rm Triangulum~Galaxy}) \approx 6.8~ {\rm kpc}\,.
\end{equation}
Turning next to an elliptical galaxy, namely,  the massive E0 galaxy NGC 1407, we have $\mathcal{D}_0 \approx 160$ kpc and $f_{DM} \approx 31$~\cite{Pot}, so that 
\begin{equation}\label{B30}
\frac{\mathcal{D}_0}{f_{DM}} \,({\rm NGC~1407}) \approx 5.2~ {\rm kpc}\,.
\end{equation}
Moreover, for the intermediate-luminosity elliptical galaxy NGC 4494, which has a half-light radius of $R_e \approx 3.77$ kpc, the dark matter fraction has been found to be $f_{DM}= 0.6 \pm 0.1$~\cite{Mor}.
Assuming that the baryonic system has a radius of $2\, R_e$, we have $\mathcal{D}_0= 4 \, R_e \approx 15$ kpc and $f_{DM} \approx 0.6$; hence,
\begin{equation}\label{B31}
\frac{\mathcal{D}_0}{f_{DM}} \,({\rm NGC~4494}) \approx 25~ {\rm kpc}\,.
\end{equation}

Let us note  that the results presented here are essesntially for the present epoch in the expansion of the universe.  Observations indicate, however, that the diameters of massive galaxies can increase with decreasing redshift $z$. For a discussion of such \emph{massive compact galaxies}, see Ref.~\cite{Per}. 

Finally,  it is interesting to consider $f_{DM}$ at the other extreme, namely,  for the case of globular star clusters and isolated dwarf galaxies. The diameter of a globular star cluster is about 40 pc. We can therefore conclude from Eq.~\eqref{B27} with $\lambda_0 \approx 3$ kpc that for globular star clusters 
\begin{equation}\label{B32}
f_{DM} \,({\rm  globular~ star~ cluster}) \lesssim  10^{-2}\,.
\end{equation}
Thus according to the virial theorem of nonlocal gravity, less than about one percent of the mass of a globular  star cluster must appear as effective dark matter if the system is sufficiently isolated and is in virial equilibrium. It is not clear to what extent such systems can be considered isolated. It is usually assumed that observational data are consistent with the existence of almost no dark matter in globular star clusters. However, a recent investigation of six Galactic globular clusters has led to the conclusion that  $f_{DM} \approx 0.4$~\cite{SBW}. The resolution of this discrepancy is beyond the scope of the present work. 

\emph{Isolated} dwarf galaxies with diameters $\mathcal{D}_0 \ll \mu_0^{-1}$ would similarly be expected to contain a relatively small percentage of effective dark matter. There is a significant discrepancy here as well, see Ref.~\cite{Oh}; again, the resolution of this difficulty is beyond the scope of this paper. In dwarf systems that  are not isolated,  the tidal influence of a much larger neighboring galaxy on the dynamics of the dwarf spheroidal galaxy cannot be ignored~\cite{Kuhn1, Kuhn2, Kuhn3}.

\section{Discussion}

Nonlocal gravity theory predicts that the amount of effective dark matter in a sufficiently isolated nearby galaxy in virial equilibrium is such that $f_{DM}$ has an upper bound, $\mathcal{D}_0 / \lambda_0$,  that is completely independent of the distribution of baryonic matter in the galaxy. However, it is possible to derive an \emph{improved} upper bound for $f_{DM}$, which does depend on how baryons are distributed within the galaxy.  To this end, we note that  Eq.~\eqref{B11} for $\mathbb{D}$ and 
$\mathbb{N}(r)<r/\lambda_0$ imply 
\begin{equation}\label{D1}
\mathbb{D} > - \frac{1}{2}\sum_{i, j}{'}~\frac{G\,m_i\,m_j}{\lambda_0}\,.
\end{equation}
If follows from this result together with Eq.~\eqref{B18} that 
\begin{equation}\label{D2}
 < \mathbb{W}_N> \, f_{DM}  > - \frac{1}{2}\sum_{i, j}{'}~\frac{G\,m_i\,m_j}{\lambda_0}\,.
\end{equation}
Let us define a characteristic length, $R_{av}$, for the average extent of the distribution of baryons in the galaxy via 
\begin{equation}\label{D3}
R_{av}\, <\mathbb{W}_N> := - \frac{1}{2}\sum_{i, j}{'}~G\,m_i\,m_j\,.
\end{equation}
Then, it follows from Eqs.~\eqref{D2} and~\eqref{D3} that
\begin{equation}\label{D4}
f_{DM} < \frac{R_{av}}{\lambda_0}\,.
\end{equation}
Clearly, $R_{av}$ depends upon the density of baryons in the galaxy. In the Newtonian gravitational potential energy in  Eq.~\eqref{D3}, 
$0< |\mathbf{x}_i-\mathbf{x}_j|  \le \mathcal{D}_0$; therefore, in general, $R_{av} \le \mathcal{D}_0$ and hence we recover from the new inequality, namely,  $f_{DM} < R_{av}/\lambda_0$, our previous less tight but more general result $f_{DM} < \mathcal{D}_0/\lambda_0$.

\begin{acknowledgments}
I am  grateful to Jeffrey Kuhn, Sohrab Rahvar and Haojing Yan for valuable discussions.  
\end{acknowledgments}

\end{document}